\documentclass{article}
\usepackage[utf8]{inputenc}
\usepackage[english]{babel}
\usepackage{amsmath,amssymb}
\usepackage{graphicx}
\usepackage{xcolor}
\usepackage{hyperref}
\usepackage{geometry}
\usepackage{authblk}
\usepackage{booktabs}
\usepackage{abstract}
\usepackage{comment}
\usepackage{csquotes}
\usepackage[normalem]{ulem}
\usepackage[backend=biber, style=numeric, sorting=none]{biblatex}
\hypersetup{colorlinks=true, linkcolor=blue, citecolor=blue, urlcolor=blue}

\usepackage{titlesec}
\titleformat{\section}{\normalfont\bfseries\large}{\thesection.}{0.5em}{}
\titlespacing{\section}{0pt}{10pt}{4pt}

\begin{document}

\title{\bfseries Gravitational Waves as a Source of Large-Scale White Noise:
New Constraints }

\author[1,2]{Gabriela Barenboim}
%\author[3,4]{Aurora Ireland}
\author[3]{Albert Stebbins}
\affil[1]{\small Instituto de F\'isica Corpuscular, CSIC--Universitat de
  Val\`encia, Paterna 46980, Spain}
\affil[2]{\small Departament de F\'isica Te\`orica, Universitat de Val\`encia,
  Burjassot 46100, Spain}
%\affil[3]{\small Department of Physics, University of Chicago, Chicago, IL 60637, USA}
%\affil[4]{\small Leinweber Institute for Theoretical Physics, Stanford University,  Stanford, CA 94305, USA}
\affil[3]{\small Fermi National Accelerator Laboratory, Theoretical Astrophysics
  Group, Batavia, IL 60510, USA}

\date{\today}
\maketitle

\begin{abstract}
\noindent
A stochastic gravitational wave (GW) background sources a shear in the flow of cosmic fluid which, through non-linear mode coupling, generates large-scale white noise (LSWN) in the kurvature density field. Building on the LSWN framework of Papers~I and II~\cite{Barenboim:2025ccc, Barenboim:2025jdg}, we derive the amplitude of this GW-induced LSWN and translate the observational non-detection of LSWN into bounds on the production redshift and density of gravity waves.  In particular, a minimal constraint on gravity waves with $z=0$ density parameter $\Omega_\mathrm{GW0}^*$ in frequency band $f_*$ generated at redshift $z_*$ must satisfy
${z_*}^2\,\Omega_\mathrm{GW0}^*<5\times10^7\,(f_*/\mathrm{nHz})^{3/2}$.  This, for example, precludes the gravity waves recently detected by pulsar timing arrays \cite{NANOGrav:2023hvm} from being present before $z_*\sim10^8$, long after the quark hadron phase transition. While orders of magnitude stronger than other constraints on gravity waves, this is a minimal LSWN constraint as realistic modeling of early universe gravity wave production, including the granularity of the gravity-wave sources and the LSWN produced by the associated acoustic waves would probably tighten this constraint by orders of magnitude.
\end{abstract}

%-------------------------------------------------------------------
\section{Introduction}
%-------------------------------------------------------------------

It was shown in~\cite{Barenboim:2025ccc} how non-linear mode coupling in the cosmic fluid on small scales inevitably generates large-scale white noise (LSWN) in density perturbations. However white noise is not observed on the largest observed scales. This non-detection of LSWN places stringent upper bounds on the amplitude of inhomogeneities on very small scales. Non-linear mode couplings that produce LSWN exist between scalar and tensor modes so small scale fluctuations in one at early times will produce large-scale fluctuations in the other at late times. This LSWN is generally a \emph{relic} of small amplitude non-linearities at early times in the sense that, before gravitational collapse, the small amplitude inhomogeneities are accurately described by linear theory.  Generically the late time super-horizon primordial power spectrum will be of the form
\begin{equation}
\Delta_\mathcal{R}^2[k]\approx
{}_\mathrm{i}\Delta_\mathcal{R}^2[k]+\frac{k_\mathrm{BH}}{k}
\label{eq:kBHdefinition}
\end{equation}
where $\mathcal{R}$ is the \emph{comoving curvature perturbation} \cite{baumann2012tasilecturesinflation}, $\Delta_\mathcal{R}^2[k]$ gives the variance in $\mathcal{R}$ per logarithmic interval in $k$, the subscript ${}_\mathrm{i}\square$ indicates its initial value and $k_\mathrm{BH}$ gives the LSWN i.e.~the amplitude generated by non-linearities at early times.  $\Delta_\mathcal{R}^2\propto k^{-1}$ corresponds to white noise in $\Delta\rho=-\nabla_\mathbf{x}^2\mathcal{R}/(4\pi\,G\,a^2)$ where $\nabla_\mathbf{x}^2$ is the comoving Laplacian.  If the horizon continues to grow then as modes $k\lesssim k_\mathrm{BH}$ enter the horizon black hole will be copiously produced.

As shown in~\cite{Barenboim:2025ccc} LSWN ($k_\mathrm{BH}>0$) can be generated from shear, $\sigma$, vorticity, $\omega$, or proper acceleration, $\dot{u}_\alpha$ of the center-of-momentum velocity.  To leading order gravity waves do not contribute to $\omega$ or $\dot{u}^\alpha$ but always result in non-zero shear in the cosmological fluid. This inevitable shear is the primary means by which one can use limits on LSWN (see eq.~\eqref{eq:DeltaRho}) - that is, on $k_\mathrm{BH}$ - to constrain the gravity wave content of the early universe. Since gravity waves do not couple to matter density inhomogeneities to any large extent\footnote{There is small coupling to the  anisotropic distribution of free-streaming cosmic neutrinos.} computing the LSWN from gravity waves is fairly straightforward. Since gravity waves are not attenuated it is also straightforward to relate measured gravity wave amplitudes to that which were present in the early universe. Gravity waves can exist primordially with correlation lengths outside the horizon at some early time or may be generated causally by the motion of matter inside the horizon.  In the latter case matter motion will produce additional LSWN leading to stronger, model dependent, constraints on sub-horizon generated gravity waves.  Here we focus on gravity waves generated inside the horizon such as would be produced by phase transitions or more exotic phenomena.

This Letter derives the GW-to-LSWN conversion quantitatively, extracts 
a bound on $\Omega_\mathrm{GW}$, and clarifies for which classes of GW 
sources this is the leading constraint. The observational constraint we use is the 99\% confidence level limit on LSWN \cite{Barenboim:2025jdg}
\begin{equation}
k_\mathrm{BH}\leq k_\mathrm{BH}^\mathrm{max}=1.80\times10^{-13}\,\mathrm{Mpc}^{-1}
\label{eq:kBHmax}
\end{equation}based on Planck 2018 data (\cite{Planck:2018vyg}).  This is the observational bound on the total LSWN from all sources which include not only GWs but also acoustic oscillations \cite{Barenboim:2025ccc}.  As the contribution from different sources are likely uncorrelated the $k_\mathrm{BH}$ contributions simply sum.  Thus the constraint of GWs should be written
$k_\mathrm{BH}^\mathrm{GW}\leq k_\mathrm{BH}^\mathrm{max}$.

Throughout we set $c=1$, work in the radiation era (RD) where $H(z)=H_0\sqrt{\Omega_r}(1+z)^2$, and use $\Delta^2(k)=4\pi\,k^3\,P(k)$.

%-------------------------------------------------------------------
\section{Curvature From Gravitational Waves}
\label{sec:GWCurvature}
%-------------------------------------------------------------------

LSWN is reflected in the curvature perturbations $\mathcal{R}$ since the kurvature density, $\Delta\rho=\nabla_\mathrm{co}^2\mathcal{R}/(4\pi\,G\,a^2)$, develops a white noise tail. The quantity $\Delta\rho$ scales like but is not the same as $\delta\rho$ in perturbation theory.  Averaging over a super-horizon volume, $\overline\square$, shear will modify the curvature perturbations by the increment
\begin{equation}
\delta\,\overline{\Delta\rho}(z)=
\frac{(1+z)^2}{2\pi G}\int_z^{z_i}dz'\,\frac{\overline{\sigma^2}(z')}{(1+z')^3}\,.
\label{eq:DeltaRho}
\end{equation}
This increment will vary randomly between different volumes leading to a white noise spectrum of $\Delta\rho$ perturbations.  This tail will dominate a roughly Harrison-Zel'dovich primordial spectrum $\sim k^1$ on the largest scales.

Now we show how to compute LSWN from gravity waves. In three steps:

\bigskip
\noindent\textbf{Step 1 --- Power Spectrum of GWs and Shear:}
A stochastic GW background with energy parameter $\Omega_\mathrm{GW}(k)$
sources a stochastic shear $\sigma$ in the cosmic fluid. For sub-horizon
modes ($k\eta \gg 1$) in the radiation era, the equal-time power spectrum is
\begin{equation}
P_{\sigma_\perp}(\eta, k)
= \frac{6\pi^2\,H(\eta)^2}{k^3}\,\Omega_\mathrm{GW}(\eta,k)\,.
\label{eq:Psigma}
\end{equation}
where $\eta$ is the conformal time. The gravity wave spectral energy content in units of the critical density is
$\Omega_\mathrm{GW}=(\frac{d}{d\ln k}\rho_\mathrm{GW})/\rho_\mathrm{crit}$,
$\rho_\mathrm{GW}=\frac{1}{32\pi\,G}\,
\langle\mathrm{tr}(\dot{\mathbf{h}}_\perp\cdot\dot{\mathbf{h}}_\perp)\rangle$, $\rho_\mathrm{crit}=\frac{3\,H^2}{8\pi\,G}$
and $\mathbf{h}_\perp$ is the dimensionless transverse strain tensor which is related to the transverse component of the shear tensor by 
$\boldsymbol{\sigma}_\perp=\frac{1}{2}\,\dot{\mathbf{h}}_\perp$.

\bigskip
\noindent\textbf{Step 2 --- Power spectrum of the squared shear:}
Gravity wave generated LSWN is proportional to 
$\sigma^2_\perp=\frac{1}{2}\,\mathrm{tr}(\boldsymbol{\sigma}_\perp\cdot\boldsymbol{\sigma}_\perp)$,
not to $\boldsymbol{\sigma}$, so the power spectrum of $\Delta\rho$ is given
by the power spectrum of $\sigma^2$.  Assuming isotropic Gaussian random noise
for $\boldsymbol{\sigma}_\perp$ the convolution formula for the power spectrum of a
quadratic field applied to eq.~\eqref{eq:Psigma} yields
\begin{equation}
P_{\sigma^2_\perp}(\eta,|\mathbf{k}|)
\simeq 72\pi^4\,H(\eta)^4
\int\frac{d^3\mathbf{p}}{(2\pi)^3}
\frac{(1+\mu^2)^2\,\Omega_\mathrm{GW}(\eta,|\mathbf{p}|)\,
                   \Omega_\mathrm{GW}(\eta,|\mathbf{k}-\mathbf{p}|)}
     {|\mathbf{p}|^3\,|\mathbf{k}-\mathbf{p}|^3}\,,
\label{eq:Psig2_Om}
\end{equation}
where 
$\mu\equiv\mathbf{p}\cdot(\mathbf{k}-\mathbf{p})/(|\mathbf{p}|\,|\mathbf{k}-\mathbf{p}|)$.
In the long-wavelength limit, $\mathbf{k}\rightarrow\mathbf{0}$ so
$|\mathbf{k}-\mathbf{p}|\to|\mathbf{p}|$, $\mu\to -1$ and $(1+\mu^2)^2\to 4$.
Writing $d^3\mathbf{p}=4\pi\,p^2\,dp$ where $p=|\mathbf{p}|$ one finds
\begin{equation}
P_{\sigma^2_\perp}(\eta,0^+)
\simeq 144\pi^2\,H(\eta)^4
\int^\infty\frac{dp}{p^4}\,\Omega_\mathrm{GW}(p,\eta)^2\,.
\label{eq:Psig2_IR}
\end{equation}
$k=0^+$ is shorthand for the limit $k\to0$.

To compute the power spectrum of $\delta\overline{\Delta\rho}$ from eq.~\eqref{eq:DeltaRho}, we need the \emph{unequal-time} correlator
$\overline{P_{\sigma^2_\perp}(\eta,\eta';\,0^+)}
=\langle\sigma^2_\perp(\eta,\mathbf{k})\,\sigma^2_\perp(\eta',-\mathbf{k})
        \rangle_{k\to0}$.
Each factor of $\sigma^2$ contributes one power of $H^2\,\Omega_\mathrm{GW}$ from eq.~\eqref{eq:Psigma}, so the equal-time result $H(\eta)^4\,\Omega_\mathrm{GW}^2$ generalizes straightforwardly to
$H(\eta)^2\,H(\eta')^2\,\Omega_\mathrm{GW}(\eta)\,\Omega_\mathrm{GW}(\eta')$
at unequal times. The remaining factor of $1/2$ arises from averaging over rapid sub-horizon oscillations of the GW modes: since each mode oscillates as $e^{\pm ik\eta}$, products of two modes at different times $\eta\neq\eta'$ contribute a phase average $\langle e^{ik(\eta-\eta')}\rangle=1/2$ after discarding the rapidly oscillating cross terms.  This gives
\begin{equation}
\overline{P_{\sigma^2_\perp}(\eta,\eta';\,0^+)}
\simeq 72\pi^2\,H(\eta)^2 H(\eta')^2
\int^\infty\frac{dp}{p^4}\,
\Omega_\mathrm{GW}(p,\eta)\,\Omega_\mathrm{GW}(p,\eta')\,.
\label{eq:Psig2_2t}
\end{equation}

\bigskip
\noindent\textbf{Step 3 --- Curvature Power Spectrum Induced by GW Shear:}
Eq.~\eqref{eq:DeltaRho} expresses $\delta\overline{\Delta\rho}$ at a single time integral of $\overline{\sigma^2_\perp}$.  The power spectrum of a quantity that is itself a time integral of a stochastic field is given by a \emph{double} time integral of the unequal-time correlator of that field,
\begin{equation}
P_{\delta\overline{\Delta\rho}}(z,0^+)
\propto \int_z dz'\int_z dz''\;
(\text{kernel})^2\;
\overline{P_{\sigma^2_\perp}(z',z'';\,0^+)}\,.
\end{equation}
where the kernel is taken from eq.~\eqref{eq:DeltaRho} and we switch from $\eta$ to $z$ for time variables. Using eq.~\eqref{eq:Psig2_2t}, one obtains
\begin{equation}
P_{\delta\overline{\Delta\rho}}(z,0^+)\simeq
\frac{18}{G^2}\,(1+z)^4\int^\infty\frac{dk}{{k}^4}\,\mathcal{I}[z,k]^2
\qquad
\mathcal{I}[z,k]=
\int_z\frac{dz'}{(1+z')^3}\,H(z')^2\,\Omega_\mathrm{GW}(z',k)
\label{eq:PdeltaDelta}
\end{equation}
if there are no other contributions to LSWN besides GWs on scales where the initial inhomogeneities are negligible. Since the power spectrum of the initial inhomogeneities goes to zero as $k\to0$ this LSWN will dominate on sufficiently large scales.  We parameterize the GW contribution to LSWN by
\begin{equation}
k_\mathrm{BH}^\mathrm{GW}\equiv\lim_{k\rightarrow0}k\,
\Delta_{\delta\mathcal{R}}^2(0,k)
=\frac{64\pi^3\,G^2}{(1+z)^4}\,P_{\delta\Delta\rho}(0,0^+)=
1152\,\pi^3\int^\infty\frac{dk}{k^4}\,\mathcal{I}[0,k]^2
\end{equation}
This is our final result for leading order LSWN generated by sub-horizon gravity waves.  This result is approximate, accurate only for gravity waves which oscillate on scales well inside the horizon which puts practical limits on the $z$ and $p$ domains of integration which is why the lower limit on $p$ and the upper limit on $z$ not stated explicitly.  We also have used the random phase approximation for these waves which is likely accurate for waves generated by numerous sources.   Next we specialize to particular scenarios of interest.

%-------------------------------------------------------------------
\section{A Minimal Model of Primordial Gravity Waves and LSWN}
\label{sec:minimal}
%-------------------------------------------------------------------

Here we develop a minimal model for the simultaneous production of gravity waves (GWs) and large-scale white noise.  It is minimal both in the sense that it is a simple construction and in that the ratio of LSWN to GW generated is smaller than other models (see \S\ref{sec:discrete}) which are more realistic representations of many early universe phenomena.

\subsection{Gaussian Minimal Model}

We are interested in gravitational waves generated deep in the radiation era where
\begin{equation}
H(z)\approx H_0\,\sqrt{\Omega_{\mathrm{r}0}}\,h(z)\,(1+z)^2\,
\qquad
h(z)\equiv\left(\frac{g_e(z)}{g_{e0}}\right)^{1/2}
\left(\frac{g_\mathrm{s}(0)}{g_\mathrm{s}(z)}\right)^{2/3}
\label{eq:Hofzwithg}
\end{equation}
where $g_\mathrm{s}(z)$ and $g_e(z)$ are the effective degrees of freedom contributing to the entropy density and energy density, respectively. In our minimal model we specialize to the case where the gravity waves are generated over a short period during the radiation era at $z\simeq z_*$.  In this case the support of the $\mathcal{I}(k)$ integral is confined to a narrow range of redshifts and we approximate $h(z)\approx h_*$, a constant.  Further approximating GW production as instantaneous,
$\Omega_\mathrm{GW}(z,k)\approx\Omega_\mathrm{GW}^*(k)\,\Theta(z_*-z)$ where $\Theta$ is the Heaviside function.  Then 
$\mathcal{I}(0,k)\approx
\frac{1}{2}\,{h_*}^2\,{H_0}^2\,{z_*}^2\,\Omega_\mathrm{r0}\,\Omega_\mathrm{GW}^*(k)$
so
\begin{equation}
k_\mathrm{BH}^\mathrm{GW}\simeq
288\,\pi^3\,
(\Omega_\mathrm{r0}\,\Omega_\mathrm{GW*})^2\,
\frac{(z_*\,h_*\,H_0)^4}{{k_*}^3}
\ .
\label{eq:kBHGWrad*}
\end{equation}
where 
\begin{equation}
\Omega_\mathrm{GW*}
\equiv\int_0^\infty\frac{dk}{k}\,\Omega_\mathrm{GW}^*(k)
\qquad
k_*\equiv\left(\int_0^\infty\frac{dk}{k^4}\,
\left(
\frac{\Omega_\mathrm{GW}^*(k)}{\Omega_\mathrm{GW*}}
\right)^2\right)^{-1/3}
\,.
\label{eq:Omegapstar}
\end{equation}
Here $k_*$ is a characteristic comoving wavenumber of GWs generated, $\Omega_\mathrm{GW*}$ is the total effective density parameter of GWs generated at $z\sim z_*$.  This last quantity is related to current density of parameter of primordial GWs by
\begin{equation}
\Omega_\mathrm{GW0}^*=\Omega_\mathrm{r0}\,\Omega_\mathrm{GW*}\,
h_*^2\,e^{-\tau_\mathrm{W}}
\label{eq:OmegaGW0}
\end{equation}
so
\begin{equation}
k_\mathrm{BH}^\mathrm{GW}\simeq
288\,\pi^3\,{\Omega_\mathrm{GW0}^*}^2\,\frac{(z_*\,H_0)^4}{{k_*}^3}\,e^{-2\,\tau_\mathrm{W}}
\ .
\label{eq:kBHGWminimal}
\end{equation}  
Here $\tau_\mathrm{W}\approx0.2$ represents damping of primordial GWs by interaction with the anisotropic stress of free streaming neutrinos (see~\cite{Weinberg:2004qa}).  What follows are rough estimates so we ignore this and the $h_*$ factor. The $z=0$ characteristic GW frequency corresponding to comoving wavenumber $k_*$ is $f_*=c\,k_*/(2\pi)$. 

\subsection{GW LSWN}

The large-scale white noise (LSWN) used so far has been white noise in the kurvature fluctuations. However, LSWN is a more universal phenomenon which we expect is present for most measurable quantities, not only the kurvature. For example the  transverse shear from GWs 
$\boldsymbol{\sigma}_\perp=\frac{1}{2}\boldsymbol{\dot{h}}_\perp$ is locally measurable.  
Here $\boldsymbol{h}_\perp$ is the transverse strain tensor commonly used to describe GWs.  One may model GWs from many phenomena as having a sharp cutoff at above some comoving wavenumber and is white noise at larger scales.  The GW density parameter at $z=0$ can then be written
\begin{equation}
\Omega_\mathrm{GW}^0(k)
=\Omega_\mathrm{GW0}\,\left(\frac{k}{k_*}\right)^{3}
\Theta(\sqrt[3]{3}\,k_* - k)\,.
\label{eq:GWmodel}
\end{equation}
where $k_*$ and $\Omega_\mathrm{GW0}$ are as defined in eq.s~\eqref{eq:Omegapstar} and \eqref{eq:OmegaGW0} so the LSWN amplitude from GWs is given by eq.~\eqref{eq:kBHGWminimal}.  For a truncated white noise spectrum most the GWs power is at wavenumbers/frequencies near the cutoff scale, $k=\sqrt[3]{3}\,k_*$, which differs only slightly from the characteristic scale, $k=k_*$.

\subsection{Leading Constraints}

If one has both a GW frequency and/or redshift in mind one can use eq.~\eqref{eq:zHorizon1} to set limits on primordial GWs through limits on $\Omega_\mathrm{GW0}^*$. This is often a more stringent constraint than previously available.  For example the  CMB $n_\mathrm{eff}$ bound, $\Omega_\mathrm{GW0}^*<1.1\times10^{-6}\,{h_{100}}^2$ \cite{Calabrese:2025ACTDR6}, for $z_*\gtrsim10^4$.  The LSWN bound is more stringent for $f_*\lesssim1\,$mHz.  

LSWN bounds become relatively much stronger when one considers higher redshifts.  For example if one considers horizon wavelength gravity waves generated during the quark hadron phase transition at temperatures $T\sim100\,$MeV with characteristic redshift $z_*\sim10^{12}$ when wavenumbers $k_*\sim1\,\mathrm{pc}^{-1}$ are entering the horizon then LSWN amplitude is
\begin{equation}
k_\mathrm{BH}^\mathrm{GW-QCD}\sim
100\,\left(\frac{\Omega_\mathrm{GW0}^\mathrm{QCD}}{10^{-8}}\right)^2\mathrm{Mpc}^{-1}
\end{equation}
which must not exceed the observational constraint of eq.~\eqref{eq:kBHmax}.  This leads to the constraint
\begin{equation}
\Omega_\mathrm{GW0}^\mathrm{QCD}<5\times10^{-16}
\qquad\text{and}\qquad
\Omega_\mathrm{GW*}^\mathrm{QCD}<8\times10^{-12}
\end{equation}
on the GW density parameter today and the GW density parameter in the radiation era.  One sees from Table~\ref{tab:QCDbounds} that the bound from non-observation of LSWN places a much more stringent constraint on primordial GWs than do other measurements.  

\begin{figure}[!t]
\centering
\includegraphics[width=\columnwidth]{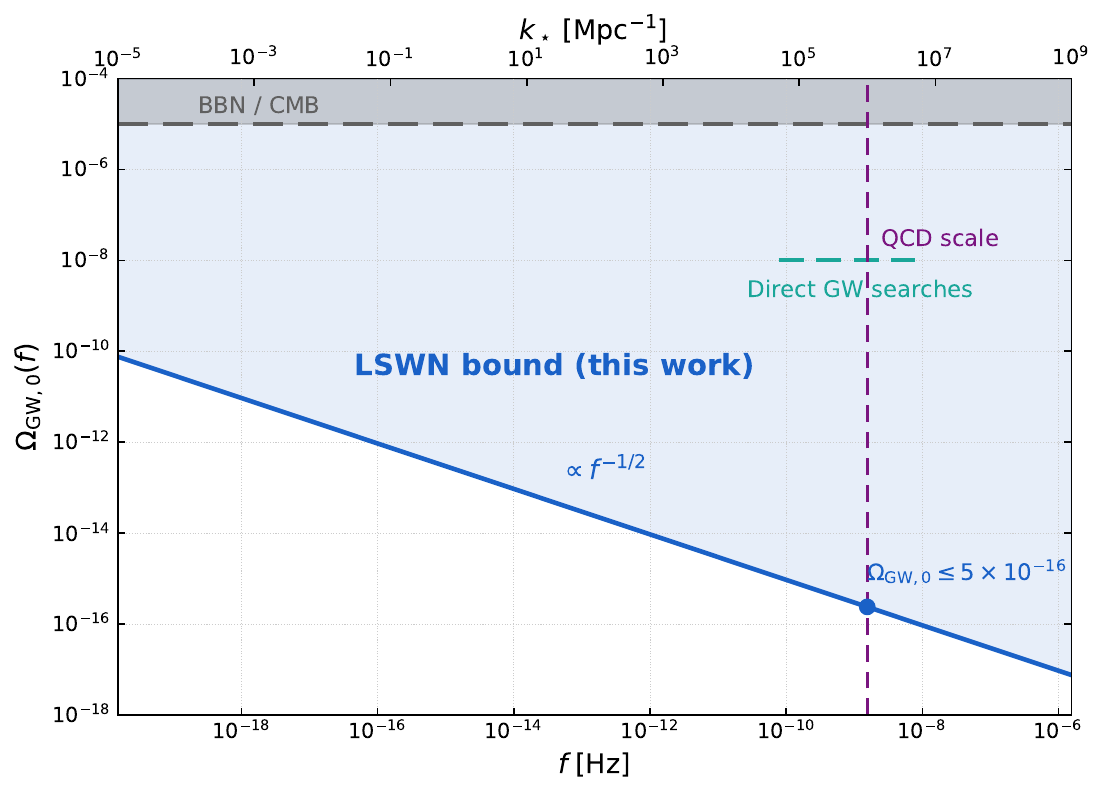}
\caption{Exclusion plot in the $(k_\star,\,\Omega_\mathrm{GW})$ plane of gravity waves produced at the horizon scale. The solid blue curve shows the LSWN bound from this work, Eq.~\eqref{eq:HorizonConstraint}, with the blue shaded region excluded. The gray dashed line and shading show the BBN/CMB/$N_\mathrm{eff}$ integrated ceiling~\cite{BBN}. The orange dashed segment shows the direct GW search bound over its relevant $k$ band~\cite{GWdirect}. The purple dashed vertical line marks the QCD horizon scale $k_\star^\mathrm{QCD}\simeq1\,\mathrm{pc}^{-1}$, where the LSWN bound gives $\Omega_\mathrm{GW0}\lesssim5\times10^{-16}$ today
($\Omega_\mathrm{GW*}\lesssim8\times10^{-12}$ at production; blue dot).
All $\Omega_\mathrm{GW}$ values on the vertical axis refer to the present-day abundance $\Omega_\mathrm{GW0}$.
Note that the blue curve bounds the GWs assuming there are no other contributions to LSWN.  This is unrealistic but, as the amount of additional LSWN is model dependent, this constraint in this figure is a conservative one.  As explained in \S\ref{sec:ConceptualFraming} more realistic modeling would likely yield a much more stringent constraint.}
\label{fig:exclusion}
\end{figure}

More generally if GWs are produced with a wavenumber a fraction $\zeta_*$ of the horizon $z_*\,k_*\simeq\,H_*/\zeta_*\simeq H_0\,\sqrt{\Omega_\mathrm{r0}}\,{z_*}^2/\zeta_*$ then the LSWN constraint in the minimal model is
\begin{equation}
k_\mathrm{BH}^\mathrm{GW}\simeq
288\,\pi^3\,{\zeta_*}^3\,\left(\frac{\Omega_\mathrm{GW0}^*}{\Omega_\mathrm{r0}}\right)^2\,k_*
\lesssim k_\mathrm{BH}^\mathrm{max}
\qquad\text{or}\qquad
\Omega_\mathrm{GW0}^*\lesssim
\frac{5\times10^{-16}}{{\zeta_*}^{3/2}}\,\sqrt{\frac{\mathrm{nHz}}{f_\mathrm{horiz}}}
\ .
\label{eq:HorizonConstraint}
\end{equation}
The power of the LSWN constraint over the broad range GW frequencies is illustrated schematically in Figure~\ref{fig:exclusion} where we take $\zeta_*=1$.  The constraint will be weakened for GWs produced well inside the horizon, $\zeta_*\ll1$, but $\zeta_*$ would have to be very small for the LSWN constraint to become weaker than constraints from direct detection of GWs.  Bounds on LSWN will therefore greatly limit the possibilities for primordial gravity wave production (see \cite{Caprini:2018mtu} for a overview of this field w/o consideration of LSWN).

\begin{table}[h]
\centering
\caption{Constraints on present-day gravity wave density parameter, $\Omega_\mathrm{GW0}$,
at wavenumber $k_*\sim1\,\mathrm{pc}^{-1}$ / frequency $f_*\sim1\,\mathrm{nHz}$
corresponding to the QCD horizon scale.  Our new bound is 7 orders of magnitude more stringent than previous bounds.
\\}
\label{tab:QCDbounds}
\renewcommand{\arraystretch}{1.3}
\begin{tabular}{lll}
\toprule
Constraint & Bound & Ref. \\
\midrule
BBN/CMB/$N_\mathrm{eff}$ (integrated) &
  $\int\Omega_\mathrm{GW}\,d\ln k \lesssim 10^{-5}$ & 
                                       \cite{Calabrese:2025ACTDR6,Yeh:2022mgl} \\
Direct GW searches &
  $\Omega_\mathrm{GW} \lesssim 10^{-8}$ & \cite{NANOGrav:2023hvm} \\
LSWN non-observation &
  $\Omega_\mathrm{GW0} \lesssim 10^{-15}$ & This work \\
\bottomrule
\end{tabular}
\end{table}

\subsection{Redshift Horizon}
\label{sec:RedshiftHorizon}

Given observational limits on LSWN: 
$k_\mathrm{BH}^\mathrm{GW}<k_\mathrm{BH}^\mathrm{max}$ 
one can, for a given population of primordial GWs characterized by $k_*$ and $\Omega_\mathrm{GW0}^*$, set an upper limit on $z_*$ which we call the \emph{redshift horizon} given by
\begin{equation}
z_*<z_\mathrm{H}^*(f_*,\Omega_\mathrm{GW0}^*)=
\frac{1}{\sqrt{6\,\Omega_\mathrm{GW0}^*}}\,
\left(\frac{c\,k_\mathrm{BH}^\mathrm{max}}{H_0}\right)^{1/4}\,
\left(\frac{f_*}{H_0}\right)^{3/4}
\simeq
2\times10^8\,\sqrt{\frac{10^{-9}}{\Omega_\mathrm{GW0}^*}}
\left(\frac{f_*}{\mathrm{nHz}}\right)^{3/4}
\label{eq:zHorizon1}
\end{equation}
which is a fundamental result of this paper.

One can apply the horizon to measured GWs or to sensitivity of existing or proposed GW telescopes.  The fiducial values used in eq.~\eqref{eq:zHorizon1} are chosen to roughly match the GWs recently detected by pulsar timing arrays (PTAs).  The maximum cosmic temperature at the  redshift horizon of the observed GWs is $\sim50\,$keV, long after big bang nucleosynthesis and during a cosmic epoch which we believe we understand.  For this reason it seems more likely the PTA detected GWs were predominantly generated at low redshifts by mundane astrophysical phenomena such as the merger of supermassive black holes.

\begin{figure}[t]
\centering
\includegraphics[width=\columnwidth]{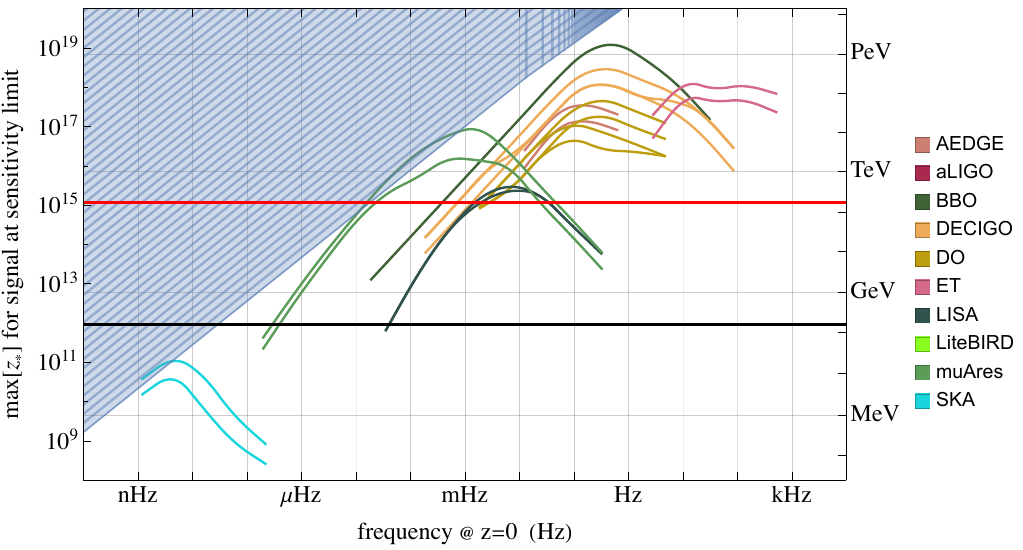}
\caption{Plotted is the redshift horizon of various surveys.  This is the maximum redshift from which various proposed survey could detect GWs.  At higher redshifts the amplitude of GWs required to exceed the sensitivity of these surveys would have produced more LSWN than allowed by observations. These curves are based on the binned $\Omega_\mathrm{GW0}$  sensitivity forecasts of ref.~\cite{2021JCAP...01..012C}.  The different colored curves are different experiments/surveys with correspondence to experiment's acronym shown in legend. There are multiple curves for all experiments except aLIGO corresponding to different model assumptions, such with and without foregrounds (which themselves are also forecasted) as explained in ref.~\cite{2021JCAP...01..012C}.  Our modeling assumes GWs are generated inside the horizon, $k_*\gg H(z_*)$. The hashed region gives parameters corresponding to generation outside the horizon, $k_*<H(z_*)$.  These $\max(z_*)$ curves need not be accurate inside of or very close to the shaded region.  
}
\label{fig:zmax}
\end{figure}

For proposed GW telescopes and ongoing surveys we do not know what GWs they will detect.  However we do have some idea as to their raw sensitivity.  From this raw sensitivity one can define the redshift horizon of the telescope/survey from eq.~\eqref{eq:zHorizon1} where we substitute the instrument's frequency band for $f_*$ and sensitivity for $\Omega_\mathrm{GW0}^*$. In fig.~\ref{fig:zmax} we plot the redshift horizon as a function of frequency using the compilation of frequency-dependent sensitivity forecasts in ref.~\cite{2021JCAP...01..012C}. We see that LSWN does not completely exclude the possibility of detecting gravity waves from the very early universe with proposed instruments, even up to cosmic temperatures above $\sim1\,$TeV for the most sensitive. 

However redshift horizons based on raw sensitivity may be overly optimistic and for a variety of reasons.  Firstly the sensitivity forecasts may themselves be optimistic; secondly because there are known low redshift astrophysical sources of GWs with expected signal well above the sensitivity limits and the true sensitivity to primordial sources after removing these "foregrounds" likely falls well short of the raw sensitivity; and thirdly because whatever mechanism that produces gravity waves in the early universe will almost certainly produce LSWN unrelated to the GWs emitted.  The redshift horizons we have used assume that the GWs are the only source of LSWN.  In the next section we will conceptually decompose LSWN into a part associated with GWs and a part which is unrelated.

%-------------------------------------------------------------------
\section{LSWN From Gravity Waves and Not From Gravity Waves}
\label{sec:LSWNnonGW}
%-------------------------------------------------------------------

In the previous section we use only the LSWN generated directly by the GWs to place bounds on any stochastic GW background generated within the horizon\footnote{The decomposition of LSWN into 
GW-induced and matter-induced contributions is derived formally 
in \S\ref{sec:GWplusKurvature}, which the reader may wish to 
consult alongside this section.}. Since GWs are generated by matter we should also consider the LSWN generated from the matter which generated the GWs.  In many cases this additional LSWN will greatly exceed the LSWN from the GWs themselves and the bounds set in the previous section greatly underestimate the total LSWN bound.  However this depends on the model dependent specifics of GW generation so it is not unreasonable to consider the bounds of the previous section as a conservative estimate.  Below we discuss how one can separate the part of LSWN which is contributed to by gravity waves.

\subsection{Conceptual Framing}
\label{sec:ConceptualFraming}

Curvature LSWN depends on the change in the kurvature, 
$K\equiv8\pi\,G\,\rho/3-\theta^2/9$ where $\rho$ is the density in the center-of-momentum (CM) frame and $\theta$ the instantaneous rate of expansion of that frame.  $K$ is a locally defined physically measurable quantity.  Conceptualize the generation of curvature as coming from a localized \emph{event} where some clump of stuff (matter + space-time curvature) is dynamically changing in a finite volume over a finite interval in time.  If the event is something like an \emph{oscillation} where the object goes back to the original state then $K$ will also go back to its original state and there will be no kurvature generation.  For such an oscillation not only does the matter go back to its original state but so does the gravitational field.  

A second conceptualization of an oscillation is that the matter goes back to its original state and so does the tidal gravitational field but what changes is that gravity waves are generated and propagate outward.  The tidal field is the \emph{near field} gravitational curvature and gravity waves (GWs) are the \emph{far field radiative} gravitational curvature producing shear, $\sigma_\perp$, of the CM frame. In this conceptualization, if the gravity waves propagate out into an expanding ($\theta>0$) \emph{cosmic medium}, they will generate kurvature and LSWN through the $\theta\,\sigma^2_\perp$ term in eq.~\eqref{eq:Kevolution}. These curvature waves will also be detectable by GW telescopes and the comparison of direct gravity wave detection to LSWN in \S\ref{sec:minimal} will be accurate.

A third conceptualization of an oscillation is that the \emph{object} that is oscillating goes back to its original state as do the tidal fields, gravity waves are generated, and in addition, this object interacts with the cosmic medium producing \emph{acoustic waves}.  The acoustic waves can contribute to all the terms which source kurvature generation in eq.~\eqref{eq:Kevolution} and will also produce LSWN.  In this 3rd conceptualization our previous comparison of direct GW detection to LSWN is missing the LSWN from the acoustic waves and thus the LSWN/GW ratio is larger and the constraints on GWs from LSWN is stronger.  In fact it can be \emph{much stronger}. The reason for this is that, while the production of GWs is fundamentally a gravitational phenomenon, acoustic waves can be generated by direct mechanical coupling of the object to the cosmic medium, i.e. GWs are $G$ or Planck suppressed while acoustic waves need not be.

As noted in ref.~\cite{Stebbins:2026ljx} kurvature generation is not fundamentally a gravitational phenomenon and can therefore be mechanical/hydrodynamical effects.  What is more important for curvature generation is the expanding medium of cosmology.  This allows shear in this medium, from either gravity or acoustic waves, to systematically increase the local kurvature and in an inhomogeneous way.  

These conceptualizations of kurvature generation probably do not capture the complexity of any realistic scenario for LSWN/GW generation, there are many other ways things can change, but we believe they provide a useful conceptual model for what is going on and what is likely to be important.  One can quantify this analysis by decomposing the LSWN into two parts LSWN$=$LWSN$_\perp$+LSWN$_\parallel$ where LWSN$_\perp$ is from the gravity waves and LSWN$_\parallel$ is from the rest and define the ratio
\begin{equation}
 \epsilon\equiv \frac{k_\mathrm{BH}^\mathrm{GW}}{k_\mathrm{BH}}.
 \label{eq:LSWNratio}
\end{equation}
As just discussed it is quite likely that $\epsilon\ll1$.  We next show how one can rigorously decompose kurvature, isolating the part that contributes to $k_\mathrm{BH}^\mathrm{GW}$.

\subsection{Longitudinal/Transverse Decomposition}
\label{sec:LongitudinalTransverse}

The central result of this section is that, in relevant scenarios, one can rigorously decompose the large-scale kurvature into a part determined by the initial conditions, a transverse part coming from transverse shear (gravity waves) and a longitudinal part from other phenomena such as matter acoustic waves.  There are additional small scale contributions to kurvature which average out on large scales.

%{\color{purple}: to leading order, the GW-induced kurvature $K_\perp$ and the matter-induced kurvature $K_\parallel$ are sourced by orthogonal components of the shear and  are therefore statistically independent. Their LSWN contributions  add in quadrature, and since 
%$\Omega_\mathrm{GW} = \epsilon\,\Omega_\mathrm{kin}$
%with $\epsilon\ll1$, the matter contribution dominates by $\epsilon^{-2}$. The remainder of this section establishes this decomposition formally.}

As we shall show one can, non-perturbatively, decompose generated LSWN into two components: one which one can think of as being generated by the matter and another which one can think of as generated by gravity waves.  As part of this decomposition, we first sketch the covariant approach to relativity \cite{Tsagas:2007yx,ellis2012relativistic}.  Decompose the space-time curvature into Ricci (both scalar and tensor) and Weyl parts.  By Einstein's equations one associates Ricci curvature with the matter.  In any frame (or "congruence" which is a 4-velocity field which fills space-time) one can decompose the Ricci curvature into density ($\rho$), pressure ($p$), energy flux ($\boldsymbol{q}$) and anisotropic stress ($\boldsymbol{\pi}$); while one can decompose the Weyl curvature into gravito-electric tensor ($\boldsymbol{E}$) and gravito-magnetic tensor ($\boldsymbol{H}$). The space-time gradient of $u^\alpha$ can be decomposed into expansion ($\theta$), vorticity ($\boldsymbol{\omega}$), shear ($\boldsymbol{\sigma}$) and proper acceleration ($\boldsymbol{\dot{u}}$).  A choice of frame allows one to unambiguously decompose all tensors into their temporal and spatial components. All these tensors referred to are purely spatial;  $\boldsymbol{q}$, $\boldsymbol{\omega}$ and $\boldsymbol{\dot{u}}$ are spatial rank 1 vectors while $\boldsymbol{\pi}$, $\boldsymbol{\sigma}$, $\boldsymbol{E}$ and $\boldsymbol{H}$ are spatial symmetric traceless rank-2 tensors also known as PSTF tensors.

All of these quantities are frame dependent but wherever normal matter is present, there is a uniquely defined center-of-momentum (CM) velocity also called the Landau-Lifshitz frame. In the CM frame then $\boldsymbol{q}=\boldsymbol{0}$ and all of the other quantities are uniquely defined and locally measurable.  This is what is meant here by \emph{covariant}.  This is in contrast to perturbation variables in other representations.  In the CM frame we can define the covariant scalar we call kurvature
\begin{equation}
\frac{8\pi\,G\,\Delta\rho}{3}=K\equiv\frac{8\pi\,G\,\rho}{3}-\frac{1}{9}\,\theta^2
\label{eq:Kdefinition}
\end{equation}
whose evolution is given exactly by
\begin{equation}
\dot{K}+\frac{2}{3}\,\theta\,K=
\frac{2}{9}\,\theta\,
\left(2\,\sigma^2-2\,\omega^2-\alpha^2-\boldsymbol{\nabla\cdot\dot{u}}\right)
-\frac{8\pi\,G}{3}\,
\mathrm{tr}\,\boldsymbol{\pi}\cdot\boldsymbol{\sigma}
\label{eq:Kevolution}
\end{equation}
where 
$\sigma^2\equiv\frac{1}{2}\,
\mathrm{tr}\,\boldsymbol{\sigma}\cdot\boldsymbol{\sigma}$ and 
$\omega^2\equiv\frac{1}{2}\,
\mathrm{tr}\,\boldsymbol{\omega}\cdot\boldsymbol{\omega}$,
$\alpha^2
\equiv\boldsymbol{\dot{u}}\cdot\boldsymbol{\dot{u}}$,
$\dot{\square}$ indicates the proper time derivative along world-lines defined by the CM frame 4-velocity field and $\boldsymbol{\nabla\cdot}\square$ is the spatial divergence.\footnote{Precise definitions of these operations can be found in \cite{Barenboim:2025ccc,Tsagas:2007yx}.}  One can express $K$ at any event as a line integral along these world-lines.  The LSWN is white noise in the power spectrum of $K$ or equivalently $\Delta\rho$.

To segregate gravity waves from other metric perturbations we would like to use a  scalar-vector-tensor (SVT) decomposition as is used in flat space or in linear perturbation theory.  Unfortunately, as shown in \S\ref{app:SVT}, beyond linear theory there is no clean way to construct a precise SVT analog when the CM frame has vorticity. However vorticity on large scales is small in our universe and we therefore believe it is a good approximation to ignore it.  As described in \S\ref{app:SVT} it is convenient to group S and V perturbations together as longitudinal modes, denoted by $\parallel$, which are complemented by the T modes which we call transverse and denote by $\perp$.  This reason for this is that, beyond linear theory, S and V modes are no longer orthogonal.  Below we use the $\parallel$/$\perp$ decomposition to separate GWs from other metric perturbations.

In the zero vorticity approximation one can separate the different contributions to kurvature:
\begin{eqnarray}
K&=&K_\mathrm{i}+K_\parallel+K_\perp+K_\times
\nonumber\\
\dot{K}_\mathrm{i}+\frac{2}{3}\,\theta\,K_\mathrm{i}&=&0
\nonumber\\
\dot{K}_\parallel+\frac{2}{3}\,\theta\,K_\parallel
&=&
\frac{2}{9}\,\theta\,
\left(2\,\sigma_\parallel^2-\alpha_\parallel^2
-\boldsymbol{\nabla\cdot\dot{u}}_\parallel\right)
-\frac{8\pi\,G}{3}\,
\mathrm{tr}\,\boldsymbol{\pi}_\parallel\cdot
             \boldsymbol{\sigma}_\parallel
\nonumber\\
\dot{K}_\perp+\frac{2}{3}\,\theta\,K_\perp&=&
\frac{4}{9}\,\theta\,\sigma_\perp^2
-\frac{8\pi\,G}{3}\,
\mathrm{tr}\,\boldsymbol{\pi_\perp\cdot\sigma_\perp}
\nonumber\\
\dot{K}_{\parallel\perp}+\frac{2}{3}\,\theta\,K_{\parallel\perp}&=&
\frac{8}{9}\,\theta\,\mathrm{tr}\,\boldsymbol{\sigma_\parallel\cdot\sigma_\perp}
-\frac{8\pi\,G}{3}\,\mathrm{tr}\,(
\boldsymbol{\pi_\perp\cdot\sigma_\parallel}+
\boldsymbol{\pi_\parallel\cdot\sigma_\perp})\ .
\label{eq:Kdecomposition}
\end{eqnarray}
In order to preserve $\boldsymbol{\omega}=0$ one requires $\boldsymbol{\dot{u}}_\perp=0$ which does not contribute to $K_\perp$. Initially $K_\parallel=K_\perp=K_{\parallel\perp}=0$ so $K_\mathrm{i}$ encodes the initial conditions. One can only measure the total kurvature, not the individual contributions, but from these formulae we may attribute the contribution of transverse (tensor) perturbations to the kurvature and hence to LSWN.  It is these transverse waves which contain the gravity wave contribution.

Longitudinal and transverse tensor are orthogonal in the sense that the contraction, one with another, spatially integrates to zero.  This is indicative that the spatial average over very large scales is small, i.e. $\overline{\boldsymbol{\sigma_\parallel\cdot\sigma_\perp}}$ and 
$\overline{\boldsymbol{\pi_\perp\cdot\sigma_\parallel}}$ are small.  Since to leading order 
$\overline{\theta\,\mathrm{tr}\,\boldsymbol{\sigma_\parallel\cdot\sigma_\perp}}\approx
3\,H\,\overline{\mathrm{tr}\,\boldsymbol{\sigma_\parallel\cdot\sigma_\perp}}$ is also small we expect that on large enough scales that $K_{\parallel\perp}$ will be smaller than both $K_\parallel$ and $K_\perp$.  That being the case the transverse (tensor) mode contribution to $K$ on large scales should be predominantly from $\dot{K}_\perp$ which to leading order is determined by
\begin{equation}
\dot{\overline{K_\perp}}+2\,H\,\overline{K_\perp}\approx
\frac{4}{3}\,H\,\overline{\sigma_\perp^2}
-\frac{8\pi\,G}{3}\,
\overline{\mathrm{tr}\,\boldsymbol{\pi_\perp\cdot\sigma_\perp}}
\end{equation}
which has solution
\begin{equation}
\overline{K_\perp}(z)=
\frac{4}{3}\,\int_z^{z_i}dz'\,\frac{(1+z)^2}{(1+z')^3}\,
\,\left(\overline{\sigma^2_\perp}(z')
+\frac{2\pi\,G\,}{H(z')}\,\overline{\mu_\perp\,\sigma^2_\perp}(z')\right)
\label{eq:KperpIntegral}
\end{equation}
where we have defined the viscosity coefficient, 
$\mu_\perp\equiv\boldsymbol{\pi_\perp\cdot\sigma_\perp}/\sigma^2_\perp$, since kinematic viscosity is merely the alignment of anisotropic stress with shear. The viscosity correction, $\sim G\,\mu_\perp/H$, is not viscosity directly but rather the gravitational effect of the viscosity, which we expect to be small. Ignoring viscosity and assuming transverse shear dominates this is identical to eq.~\eqref{eq:DeltaRho} when making the identification $\overline{K_\perp}=8\pi\,G\,\delta\overline{\Delta\rho}/3$.

Since $\boldsymbol{\sigma}_\parallel$ and $\boldsymbol{\sigma}_\perp$ are orthogonal, their contributions to $\overline{\sigma^2}$, and hence to $K$ via eq.~\eqref{eq:Kevolution}, are additive. The transverse piece $K_\perp$ is identified with GW-induced kurvature and scales as $\Omega_\mathrm{GW}^2$.  Note that $K_\parallel$ and $K_\perp$ are \emph{not} orthogonal functions.  The ratio of eq.~\eqref{eq:LSWNratio} is
\begin{equation}
\epsilon\simeq\frac{\mathrm{var}\,\overline{K_\perp}}
{\mathrm{var}\,(\overline{K_\parallel}+\overline{K_\perp})}
\label{eq:LSWNratio2}
\end{equation}
which might be a very small number.

\section{Simultaneous Gravity Wave and Kurvature Generation}
\label{sec:GWplusKurvature}

In this section we confine our attention to the close connection between the generation of gravity waves and kurvature LSWN in the context of weak-field gravity. A generalization of the statement in \S\ref{sec:GWCurvature} that "gravity waves generate kurvature" ($K_\perp$ in particular) is that "transverse shear ($\boldsymbol{\sigma}_\perp$) generates kurvature".  Gravity waves are generally described by the transverse strain $\boldsymbol{h}_\perp$ and since $\boldsymbol{\sigma}_\perp=\frac{1}{2}\boldsymbol{\dot{h}}_\perp$ for the moment we do not distinguish gravity waves and $\boldsymbol{\sigma}_\perp$.  In \S\ref{app:TensorGeneration} it was shown that to leading order $\boldsymbol{\sigma}_\perp$ is generated by transverse anisotropic stress, $\boldsymbol{\pi}_\perp$ and since $\boldsymbol{\sigma}_\perp$ generates $K_\perp$ one can say that transverse anisotropic stress simultaneously generates gravity waves and kurvature.

\subsection{Discrete Event Model}
\label{sec:discrete}

To model the simultaneous generation of GWs and $K_\perp$ in the early universe one needs to model $\boldsymbol{\pi}_\perp$, i.e.~how the matter behaves.  One can however relate two late time manifestations of early universe $\boldsymbol{\pi}_\perp$: $\Omega_\mathrm{GW}^\mathrm{eu}$ (current density parameter of gravity waves generated in the  early universe) and $k_\mathrm{BH}^\perp$ (LSWN generated by $K_\perp$) without specifying much about the $\boldsymbol{\pi}_\perp$ which generated them.  Here we also assume that the viscosity term is negligible, $G\,\mu_\perp/H\ll1$.

One can simply relate gravity waves and kurvature generated by a population of localized discrete events in the early universe.  They are discrete in the sense that the gravity waves produced from different objects do not systematically interfere with each other.  One can achieve discreteness by placing the events at random locations which we will assume here.

To characterize the events we need only specify the " effective energy" of gravity waves produced by event $i$ as a function of redshift $\mathcal{E}_i(z)$ including redshifting of the energy.  While strictly speaking gravity waves do not possess energy the effective energy density of a rapidly oscillating gravity wave, averaged over many oscillations, is 
$\rho_\mathrm{GW}
=\tfrac{1}{32\pi\,G}\mathrm{tr}\boldsymbol{\,\dot{h}_\perp\cdot\dot{h}_\perp}
=\tfrac{1}{4\pi\,G}\,\sigma_\perp^2$.  Here we take this as the definition of "tensor energy" whether or not the waves are rapidly oscillating so that
$\mathcal{E}_i(z)=\tfrac{1}{4\pi\,G}\int\tfrac{d^3\mathbf{x}}{(1+z)^3}\,
\sigma_{\perp i}^2(z,\mathbf{x})$
where $\mathbf{x}$ is the comoving coordinates scales to $z=0$. Since after the event $\mathcal{E}_i\propto(1+z)$ it is convenient to reparameterize in terms of a quantity which is constant.  Here we choose the comoving volume containing this much energy:
\begin{equation}
\mathcal{V}_i(z)\equiv(1+z)^3\,\frac{\mathcal{E}_i(z)}{\bar{\rho}(z)}
\qquad
\mathcal{E}_i(z)=\tfrac{2}{3}\int d^3\mathbf{x}\,
\frac{\sigma_{\perp i}^2(z,\mathbf{x})}{H(z)^2} \ .
\end{equation}
which is constant after GW generation during the radiation era, i.e. $\mathcal{V}_i(z)\rightarrow\mathcal{V}^*_i$. If the event population has comoving number density $n_*$ then the GWs generated by these events will today have density parameter
\begin{equation}
\Omega_\mathrm{GW0}^*=n_*\,\langle\mathcal{V}^*_i\rangle_*\,\Omega_\mathrm{r0}
\label{eq:OmegaGWstar}
\end{equation}
where $\langle\square_i\rangle_*$ gives the event weighted mean of $\square_\mathrm{i}$.

Transverse anisotropic stress from each event $i$ will also produce kurvature, $K_{\perp i}$.  Integrating $K_{\perp i}$ over volume uses the same volume integral of squared transverse shear used to determine $\mathcal{V}_i$. After GW generation ends the eq.~\eqref{eq:KperpIntegral} redshift integral converges rapidly so asymptotically we can define a quantity with dimensions of comoving length for each event
\begin{equation}
\mathcal{L}_i\equiv\lim_{z\rightarrow0}
\int d^3\mathbf{x}\,\frac{K_{\perp i}(z)}{(1+z)^2}
\approx
\frac{4}{3}\,\int_0^\infty\frac{ dz'}{(1+z')^3}\,
\int d^3\mathbf{x}\,\sigma^2_\perp(z',\mathbf{x})
=2\,\int_0^\infty\frac{ dz'}{(1+z')^3}\,\mathcal{V}_i(z')\,H(z')^2\ .
\end{equation}
Unlike the GWs which disperse at the speed of light for localized events the kurvature remains localized at roughly the horizon scale when the event occurred. On comoving scales larger than this scale the power spectrum of the transverse kurvature produced by all events, $K_\perp=\sum K_{\perp i}$ will have a white noise spectrum.  The amplitude of this contribution to LSWN is parameterized by 
\begin{equation}
k_\mathrm{BH}^*=\frac{9}{8\pi^2}\,n_*\,\langle \mathcal{L}_i^2\rangle_* \ .
\end{equation}
Using eq.~\eqref{eq:Hofzwithg} for the radiation era and defining $z_i^*$ by
\begin{equation}
(1+z_i^*)^2\equiv2\int\,dz'\,(1+z')\,h(z)^2\,\frac{\mathcal{V}_i(z')}{\mathcal{V}_i^*}
\end{equation}
we find
\begin{equation}
\mathcal{L}_i=\Omega_\mathrm{r0}\,H_0^2\,(1+z_i^*)^2\,\mathcal{V}_i^*
\end{equation}
so
\begin{equation}
k_\mathrm{BH}^*
=\frac{9}{8\pi^2}\,\Omega_\mathrm{r0}^2\,H_0^4\,n_*\,
\langle (1+z_i^*)^4\,\mathcal{V}_i^{*2}\rangle_*
\label{eq:kBHstar}
\end{equation}
The quantity $z_i^*$ is defined so that it gives the correct values for $\mathcal{L}_i$ and $\mathcal{V}_i^*$ if GWs were all produced at $z=z_i^*$ (ignoring small $h(z)$ corrections). This characteristic redshift corresponds to $z_*$ in the fiducial model of \S\ref{sec:minimal} but here $z_i^*$ may vary from event to event. This characteristic redshift depends on the time profile of transverse shear squared. One generally expects $\mathcal{V}_i(z)$ to increase monotonically unless there is systematic destructive interference of GWs.  If the turn on is rapid then $z_i^*$ is when most GWs were generated and kurvature growth follows within a few Hubble times.

Eq.s~\eqref{eq:OmegaGWstar} and \eqref{eq:kBHstar} give expressions for two low redshift observables\footnote{These "observables" may be mixed with other indistinguishable contributions to GWs and LSWN.}, $\Omega_\mathrm{GW}^*$ and $k_\mathrm{BH}^*$(eq.~\eqref{eq:OmegaGWstar}) as a function of three parameters characterizing their generation: the comoving number density of events ($N_*$), the GW output of these events, $\mathcal{V}_i^*$, and a characteristic redshift of GW emission $z_i^*$.  The frequencies of the GWs emitted or observed play no role in determining the two parameters, that information is removed when one squares the shear.  Measurements of GW frequencies would further refine modeling of early universe generation of GWs and kurvature beyond the crude characterization in the event model. 

The parameters $z_i^*$ represent the characteristic redshifts of each event separately.  A characteristic redshift of the population of events is defined by
\begin{equation}
(1+z_*)^4=\frac{\langle (1+z_i^*)^4\,\mathcal{V}_i^{*2}\rangle_*}
                             {\langle\mathcal{V}_i^{*2}\rangle_*}
\end{equation}
and a more intuitive parameterization of the density of events is the number of events per Hubble volume at $z=z_*$:
\begin{equation}
N_*\equiv\frac{h(z_*)^3\,(1+z_*)^3\,n_*}{H(z_*)^3}
=\frac{n_*}{{H_0}^3\,{\Omega_{\mathrm{r}0}}^{3/2}\,(1+z_*)^3} \ .
\end{equation}
If there is much fewer than one event per horizon volume, $N_*\ll1$, the events may be considered \emph{rare} while if there are much greater than one event, $N_*\gg1$ one may consider them \emph{frequent}.  One can relate the two observables 
\begin{equation}
k_\mathrm{BH}^*
=\frac{9}{8\pi^2}\,
\frac{\langle\mathcal{V}_i^{*2}\rangle_*}{{\langle\mathcal{V}^*_i\rangle_*}^2}\,
\frac{(1+z_*)\,H_0}{N_*}\,
\frac{{\Omega_\mathrm{GW0}^*}^2}{\Omega_\mathrm{r0}^{3/2}}\ .
\label{eq:kBHofOmegaGW}
\end{equation}
Upper limits on LSWN place upper limits on the amount of GWs ($\Omega_\mathrm{GW}^*$) and the redshift at which it could be generated ($z_*$) and lower limits on the density of events which generated it.  A smaller number of events leads to a more "grainy" kurvature field and thus more LSWN.  For the empirical upper limit on LSWN and a measurements of the GW energy density one can set bounds on the parameters of early generation of these GWs
\begin{equation}
\frac{1+z_*}{N_*}<\frac{8\pi^2}{9}\,
\frac{{\langle\mathcal{V}^*_i\rangle_*}^2}{\langle\mathcal{V}_i^{*2}\rangle_*}\,
\frac{{\Omega_\mathrm{r0}}^{3/2}}{{\Omega_\mathrm{GW0}^*}^2}\,
\frac{k_\mathrm{BH}^\mathrm{max}}{H_0}
=62\,
\left(\frac{10^{-8}}{\Omega_\mathrm{GW0}^{*}}\right)^2\,
\frac{k_\mathrm{BH}^\mathrm{max}}{1.8\times10^{-13}\,\mathrm{Mpc}^{-1}}
\end{equation}
where $10^{-8}$ gives the density parameter of GWs detected by PTAs and we have used 
${\langle\mathcal{V}_i^{*2}\rangle_*\ge\langle\mathcal{V}^*_i\rangle_*}^2$.
We see that these GWs could only have been generated by events at fairly low redshifts unless the number of events per horizon volume was large. $N_*\gtrsim10^8$ for the quark hadron phase transition at $z_*\sim10^{12}$. 

Note that the $N_*\sim1$ bounds on $z_*$ in the discrete event model for GW generation are much more stringent than that for the minimal model of \S\ref{sec:minimal}.  A major difference between the two  models is that the minimal model assumes homogeneous Gaussian random for the GWs whereas in the event model the GWs do not have random phases and are in more of a coherent state assumed to be coming from a few emission events.  Since LSWN is driven by non-linear terms in the shear it is much more sensitive to non-Gaussianities than the power spectrum of GWs.  The effects of non-Gaussianity are modulated by the parameter $N_*$.  In the absence of a GW emission mechanism conservative bounds on the redshift of origin of primordial GWs should use the Gaussian not the event model of this section however in some cases the event model may be more accurate.

%-------------------------------------------------------------------
\section{Implications for Early Universe Physics}
%-------------------------------------------------------------------

The LSWN constraints given in this paper have important consequences for cosmological phase transitions and other phenomena in the early universe. Consider the QCD transition at $T_\mathrm{QCD}\sim100\,\mathrm{MeV}$. If it were strongly first-order, it would involve significant bulk kinetic energy $\Omega_\mathrm{kin}$ in bubble collisions and turbulent flows, which would in turn radiate a fraction of the free kinetic energy into a GW background with $\Omega_\mathrm{GW} = \varepsilon_\mathrm{GW}\,\Omega_\mathrm{kin}$ with $\varepsilon_\mathrm{GW}\ll1$ (see \cite{Caprini:2018mtu}).  The LSWN non-observation bound constrains $\Omega_\mathrm{GW}$ via
Eq.~\eqref{eq:kBHGWminimal}. Since there is a direct mechanical coupling of the $\Omega_\mathrm{kin}$ to the cosmic medium a phase transition is liable to radiate a much larger fraction of its energy to acoustic waves,
$\Omega_\mathrm{aw} = \varepsilon_\mathrm{aw}\,\Omega_\mathrm{kin}$ with 
$\varepsilon_\mathrm{aw}\gg\varepsilon_\mathrm{GW}$.  These acoustic waves will generate much more LSWN than the GWs (see \S\ref{sec:ConceptualFraming}) leading effectively to a much tighter LSWN constraint on $\Omega_\mathrm{GW}$ from eq.~\eqref{eq:kBHGWminimal}. 

To compare kurvature from gravity and acoustic waves note that for short wavelength waves their effective time averaged energy density during the radiation era is 
$\rho_\mathrm{GW}=\frac{1}{4\pi\,G}\sigma_\perp^2$ and 
$\rho_\mathrm{ac}=\frac{3}{2\pi\,G}\,(H/k_\mathrm{phys})^2\sigma_\parallel^2$ where $k_\mathrm{phys}$ is the physical wavenumber and $H$ the Hubble parameter. Using $\alpha^2_\parallel=\sigma^2_\parallel$ the volume averaged kurvature evolution (eq.~\eqref{eq:Kevolution}) written in terms of $\Delta\rho\equiv3\,K/(8\pi\,G)$ is given by
\begin{equation}
\dot{\overline{\Delta\rho}}+2\,H\,\overline{\Delta\rho}=
2\,H\,\left(\overline{\rho}_\mathrm{GW}
            +\frac{1}{12}\,\frac{{k_\mathrm{phys}}^2}{H^2}\,\overline{\rho}_\mathrm{aw}\right)
            \,.
\label{eq:DeltaRhoEvolution}
\end{equation}
Thus we see that not only is it likely that 
$\overline{\rho}_\mathrm{GW}\ll\overline{\rho}_\mathrm{aw}$
but if these waves are sub-horizon the acoustic waves are much more effective in generating curvature than the gravity waves.  If the acoustic and gravity waves are spatially correlated one can estimate the fraction of LSWN generated by GWs, eq.s~\eqref{eq:LSWNratio}~and~\eqref{eq:LSWNratio2} ,
\begin{equation}
\epsilon\sim\frac{{\varepsilon_\mathrm{GW}}^2}
{({\varepsilon_\mathrm{GW}}+\frac{1}{12}\,(k_\mathrm{aw}\,\eta_\mathrm{aw})^2\,
{\varepsilon_\mathrm{aw}})^2}
\label{eq:LSWNratio3}
\end{equation}
where $k_\mathrm{aw}$ and $\eta_\mathrm{aw}$ are characteristic comoving wavenumbers and conformal time of acoustic wave emission.  This is likely to be an exceedingly small number meaning that LSWN is probably predominantly from acoustic waves and not from gravity waves.

The PTA measured GWs (see \cite{NANOGrav:2023hvm}) could not possibly have come from such a high redshift as the QCD transition because it would produce observable LSWN.  This bound on
QCD phase transition is far more stringent than direct GW constraints,
effectively ruling out a strongly first-order transition with any significant bulk
kinetic energy fraction.  The same is true for other phase transitions or any other phenomena at high redshifts which generate significant bulk flows.  Thus, the early universe must have been very quiet.

The early universe may be quiet, but perhaps we could still hear it if we had sensitive enough GW telescopes.  Figure~\ref{fig:zmax} suggests that this may just be possible under very optimistic assumptions.  However, for reasons elucidated in \S\ref{sec:RedshiftHorizon}, and particularly because the redshift horizon should be decreased by a factor $\sim\epsilon$ it seems unlikely that any gravity wave signal consistent with the LSWN would be detectable by future GW telescopes except if from fairly low redshifts.  Detailed models of gravity and acoustic wave production are required to make realistic estimates of redshift horizons as well as LSWN constraints on early universe physics.

%The logic is simply this: if even the subdominant GW-induced LSWN is below the observational threshold, the dominant matter LSWN --- larger by $\epsilon^{-2}$ --- would have been detected long ago had the kinetic energy been large. Its non-detection closes the case.

%-------------------------------------------------------------------
\section{Summary}
%-------------------------------------------------------------------

We have computed the large-scale white noise power spectrum sourced by a
stochastic gravitational wave background, and translated the LSWN
non-observation bound of Ref.~\cite{Barenboim:2025jdg} into constraints on the GW
energy density. For GWs produced inside the horizon during radiation domination the minimal model for the GW-induced LSWN amplitude is given by eq.~\eqref{eq:kBHGWrad*}:
$k_\mathrm{BH}^\mathrm{GW}
\simeq 36\,{\Omega_\mathrm{GW0}^*}^2\,(z_*\,H_0)^4/{f_*}^3$ where $\Omega_\mathrm{GW0}^*$ is the density parameter of the GWs produced in the early universe, $z_*$ is the characteristic redshift when they are produced and $f_*$ is their $z=0$ frequency.  This assumes Gaussian initial conditions for the gravity waves. The constraint would be tighter if sources of gravity waves were localized, giving greater granularity to the kurvature field (see \S\ref{sec:discrete}). Furthermore, the LSWN produced by GWs must be added to LSWN produced by the matter which produced the GWs.  This is likely to increase the LSWN in this paper by orders of magnitude (see \S\ref{sec:ConceptualFraming}) making the LSWN bound on GWs orders of magnitude more stringent.  One must also include LSWN from acoustic waves unassociated with GW production.  Detailed modeling of specific scenarios for early universe gravity wave production will be required to obtain these more stringent LSWN bounds.

An exact GR formalism for isolating the gravity wave contribution to kurvature inhomogeneities is given in \S\ref{sec:LongitudinalTransverse}\,\&\,\ref{app:SVT} which can be useful for more detailed modeling of GWs, especially if generated in a strong gravitational field environment.

\bigskip

\section*{Acknowledgments}
\noindent
GB is supported by the Spanish grants  PID2020-113775GB-I00 
(AEI/10.13039/501100011033), and by the European ITN project HIDDeN (H2020-MSCA-ITN-2019/860881-HIDDeN). 
%A.I.\ is supported by the Leinweber Institute for Theoretical Physics.
This work was produced by Fermi Forward Discovery Group, LLC under Contract No. 89243024CSC000002 with the U.S. Department of Energy, Office of Science, Office of High Energy Physics. Publisher acknowledges the U.S. Government license to provide public access under the DOE Public Access Plan (www.energy.gov/doe-public-access-plan).

%-------------------------------------------------------------------

\bigskip

\begin{appendix}
%%%%%%%%%%%%%%%%%%%%%%%%%%%%%%%%%%%%%%%%%%%%%%%%%%%%%%%%%%%%
%%%%%%%%%%%%%%%%%%%%%%%%%%%%%%%%%%%%%%%%%%%%%%%%%%%%%%%%%%%%

\section{SVT, Vorticity, Gravity Waves}
\label{app:SVT}

The results in this paper are based upon formulae from an exact covariant representation of GR e.g.~eq.~\eqref{eq:Kevolution}.  Since the focus of this paper is on gravity waves it will be instructive to define what we mean by gravity waves in the same exact covariant representation which is what is attempted in this Appendix. In unperturbed FLRW space-times and more generally in FLRW linear perturbation theory one usually decomposes the metric perturbation into scalar (S), vector (V) and tensor (T) components.  This SVT decomposition can be generalized, with some limitations, to arbitrary time slicing of space-time as we now describe.  A non-zero center-of-momentum vorticity is an obstacle to many of the convenient features of the homogeneous space-time SVT decomposition. The vorticity-dependent modification of SVT, will, in many circumstances, be small.  

\subsection{Generalized SVT and \texorpdfstring{$\parallel\perp$}{|| and perpendicular} Decompositions}

For any 3-D manifold any vector field can be uniquely Hodge decomposed into a longitudinal (scalar or S), transverse (vector or V) and possibly harmonic vector  fields~\cite{Frankel:2011}.:
\begin{equation}
\boldsymbol{v}=\boldsymbol{v}_\mathrm{S}+\boldsymbol{v}_\mathrm{V}+\boldsymbol{v}_\mathrm{H}
\qquad 
\boldsymbol{v}_\mathrm{S}=\boldsymbol{\nabla}s
\qquad
\boldsymbol{\nabla\cdot v}_\mathrm{V}=0
\qquad
\nabla^2\boldsymbol{v}_\mathrm{H}=0
\end{equation}
The existence of a non-zero harmonic field depends on the topology and/or boundary conditions. In cosmological applications the harmonic component either does not exist or is neglected.  On any compact manifold the longitudinal and transverse fields are orthogonal in the sense that
\begin{equation}
\int d^3V\,\boldsymbol{v}_\mathrm{S}\cdot\boldsymbol{u}_\mathrm{V}=0
\end{equation}
where $d^3V$ is the volume element from the metric manifold's $\boldsymbol{g}$.

A 3D symmetric traceless rank 2 tensor field  may also be York+Hodge decomposed into scalar (S), vector (V), tensor (T) and harmonic (H) fields~\cite{York:1973,Gourgoulhon:2007}.
\begin{equation}
\boldsymbol{M}=\boldsymbol{M}_\mathrm{S}+\boldsymbol{M}_\mathrm{V}+\boldsymbol{M}_\mathrm{T}+\boldsymbol{M}_\mathrm{H}
\qquad 
\boldsymbol{M}_\mathrm{S}=\boldsymbol{\nabla}\boldsymbol{\nabla}s
-\frac{1}{3}\,\boldsymbol{g}\,\nabla^2s
\qquad
\boldsymbol{M}_\mathrm{V}=
\boldsymbol{\nabla w}+(\boldsymbol{\nabla w})^\mathrm{T} \qquad \boldsymbol{\nabla\cdot w}=0
\end{equation}
and again we neglect any harmonic component, $\boldsymbol{M}_\mathrm{H}$.  Unless the spatial manifold has constant curvature S and V tensors are not, in general, orthogonal:
\begin{equation}
\int d^3V\,\mathrm{tr}\,\boldsymbol{M}_\mathrm{S}\cdot\boldsymbol{N}_\mathrm{V}\ne0\ .
\end{equation}
but both S and V tensors are always orthogonal to T tensors:
\begin{equation}
 \int d^3V\,\mathrm{tr}\,\boldsymbol{M}_\mathrm{S}\cdot\boldsymbol{N}_\mathrm{T}
=\int d^3V\,\mathrm{tr}\,\boldsymbol{M}_\mathrm{V}\cdot\boldsymbol{N}_\mathrm{T}=0\ .
\end{equation}
Thus the SVT decomposition is not as useful on an arbitrary spatial manifolds as it is in the constant curvature spatial manifolds we are used to.  For this reason we will not distinguish S and V denoting linear combinations of S and V tensors as \emph{longitudinal} tensors (denoted by $\parallel\equiv\mathrm{S}\cup\mathrm{V}$) and denoting the T tensors as \emph{transverse} tensors (denoted by $\perp\equiv\mathrm{T}$). $\parallel$ tensors are orthogonal to $\perp$ tensors~\cite{York:1973}.  Either the SVT or $\parallel\perp$ decomposition serves our purpose of separating out the transverse traceless (T or $\perp$) tensors.  Note however in space-time both decompositions depend on the time-slicing and are separately defined for each time-slice.  Even parallel transport can mix S, V and T.

\subsection{Neglecting Vorticity}

In \S\ref{sec:LSWNnonGW} we defined "spatial" tensors as those perpendicular to a 4-velocity "frame" (congruence).  This is \emph{not} the same as a tensor being in the tangent space of a particular time slice.  In fact, when the 4-velocity field has non-zero vorticity, there is no time slice in whose tangent space these spatial vectors lie. Thus the nice properties of the SVT/$\parallel\perp$ decompositions do not apply directly to the spatial tensors defined in \S\ref{sec:LSWNnonGW} unless one can neglect vorticity.  One can, of course, always choose a frame which has no vorticity but this arbitrariness of choice, akin to a choice of coordinates (gauge), is something we want to avoid. Instead we prefer the uniquely defined center-of-momentum (CM) frame. While there are, no doubt, deterministic prescriptions for defining vorticity-free rest frame, we will instead just assume vorticity is negligible.  This may not be a bad approximation for modeling our universe on large scales.

In \S\ref{sec:LSWNnonGW} we have described space-time geometry in terms of 1) the center-of-momentum (CM) frame; 2) the CM frame shear ($\boldsymbol{\sigma}$), vorticity ($\boldsymbol{\omega}$), proper acceleration ($\boldsymbol{\dot{u}}$); 3) the total density ($\rho$), pressure ($p$) and anisotropic stress $\boldsymbol{\pi}$ in the CM frame; and 4) the gravito-electric ($\boldsymbol{E}$) and gravito-magnetic ($\boldsymbol{H}$) field.  All tensor quantities are purely spatial and $\boldsymbol{\sigma}$, $\boldsymbol{\pi}$, $\boldsymbol{E}$ and $\boldsymbol{H}$ are PSTF (spatial symmetric traceless rank 2 tensors. Assuming $\boldsymbol{\omega}=0$ one can slice time into a 3D manifolds which are spatial in the CM frame.  Thus all the spatial tensors are also tensors on these orthogonal time slices and may thus be SVT decomposed.  Keeping $\boldsymbol{\omega}=0$ on all time-slices constrains the other tensors and in particular $\boldsymbol{\dot{u}}_\mathrm{V}=\boldsymbol{\sigma}_\mathrm{V}
=\boldsymbol{H}_\mathrm{V}=0$.  To 1st order in perturbation from an FLRW space-time 
$\boldsymbol{\pi}_\mathrm{V},\,\boldsymbol{E}_\mathrm{V}\approx0$.

\subsection{Tensor Modes in GEM}

Gravito-electromagnetism (GEM) is the theory of the Weyl curvature tensor expressed in terms of $\boldsymbol{E}$ and $\boldsymbol{H}$ and is described by\footnote{
$\dot{\square}$, $\mathrm{div}\square$ and $\langle\boldsymbol{\nabla\times\square}\rangle$ are precisely defined in \cite{Tsagas:2007yx} and mean roughly what you think they mean.  Here $\boldsymbol{\nabla}$ is a spatial gradient and $\langle\square\rangle$ projects rank 2 tensors, $\square$, into their traceless spatial symmetric (PSTF) part.}
\begin{align}
&\dot{\boldsymbol{E}}+\theta\,\boldsymbol{E}-\langle\boldsymbol{\nabla\times H}\rangle=
8\pi\,G\,\boldsymbol{J}_\mathrm{E}
&\boldsymbol{\nabla\cdot E}=8\pi\,G\,\boldsymbol{\rho}_\mathrm{E}
\nonumber\\
&\dot{\boldsymbol{H}}+\theta\,\boldsymbol{H}-\langle\boldsymbol{\nabla\times E}\rangle
=8\pi\,G\,\boldsymbol{J}_\mathrm{H}
&\boldsymbol{\nabla\cdot H}=8\pi\,G\,\boldsymbol{\rho}_\mathrm{H}
\label{eq:GEM}
\end{align}
which has a same form as Maxwell's equations except for the $\theta$ terms which gives field dilution when expands ($\theta>0$) and field concentration when matter contracts ($\theta<0$).  $\rho_\mathrm{E}$ and $\rho_\mathrm{H}$ are the gravito-electric and magnetic charge densities while
$\boldsymbol{J}_\mathrm{E}$ and $\boldsymbol{J}_\mathrm{H}$ are the gravito-electric and magnetic current density.  A major difference from Maxwell is that $\rho_\mathrm{E}$ and $\rho_\mathrm{H}$ are spatial vectors not scalars and $\boldsymbol{J}_\mathrm{E}$ and $\boldsymbol{J}_\mathrm{H}$ are PSTF tensors not vectors.  The charge and current densities are a sum of terms linear and quadratic in $\theta$, $\boldsymbol{\sigma}$, $\boldsymbol{\pi}$, $\boldsymbol{E}$, $\boldsymbol{H}$, $\boldsymbol{\omega}$ and $\boldsymbol{q}$. The gravito-magnetic field, $\boldsymbol{H}$ is not a separate degree of freedom since it can be derived from other quantities on a spatial hypersurface:
$\boldsymbol{H}=\langle\boldsymbol{\nabla\times\sigma}\rangle
+\langle\boldsymbol{\nabla\omega}\rangle
+2\,\langle\boldsymbol{\dot{u}\,\omega}\rangle$. The use of the curl of a tensor does provide some elegance to GEM and allows one to express certain identities in the Maxwell's equations form.

In the zero vorticity case $\boldsymbol{H}=\langle\boldsymbol{\nabla\times\sigma}\rangle$ and to linear order $\boldsymbol{H}_\mathrm{S}=\boldsymbol{H}_\mathrm{V}\approx0$ so $\boldsymbol{H}\approx\boldsymbol{H}_\mathrm{T}$.  In linear theory, with or without vorticity, the tensor component of the shear is half the proper time derivative of the dimensionless transverse traceless strain tensor i.e.~$\boldsymbol{\sigma}_\mathrm{T}=\frac{1}{2}\boldsymbol{\dot{h}}_\mathrm{T}$.  Thus in the linear theory $\boldsymbol{\sigma}_\mathrm{T}$ or $\boldsymbol{H}$ are different representations of tensor metric perturbations as normally expressed.  

Here we consider only the CM frame ($\boldsymbol{q}=0$) with zero vorticity ($\boldsymbol{\omega}=0$) where $\boldsymbol{\sigma}$ determines $\boldsymbol{H}$ and the dynamical part of GEM may be rewritten
\begin{eqnarray}
\dot{\boldsymbol{E}}+\theta\,\boldsymbol{E}&=&
\nabla^2\boldsymbol{\sigma}-8\pi\,G\,\boldsymbol{J}_\mathrm{E}
\nonumber\\
\dot{\boldsymbol{\sigma}}+\frac{2}{3}\,\theta\,\boldsymbol{\sigma}&=&
4\pi\,G\,\boldsymbol{\pi}-\boldsymbol{E}
+\boldsymbol{\langle\sigma\cdot\sigma\rangle}
+\boldsymbol{\langle\nabla\dot{u}\rangle}
+\boldsymbol{\langle\dot{u}\,\dot{u}\rangle} \ .
\end{eqnarray}
In the irrotational CM frame to 1st order in perturbation from flat FLRW
\begin{align}
&\boldsymbol{J}_\mathrm{E}\approx
-\frac{1}{2}\,(\rho+p)\,\boldsymbol{\sigma}
-\frac{1}{2}\,\boldsymbol{\dot{\pi}}
-\frac{1}{6}\,\theta\,\boldsymbol{\pi}
&\boldsymbol{\rho}_\mathrm{E}\approx
 \frac{1}{3}\,\boldsymbol{\nabla}\rho
\nonumber\\
&\boldsymbol{J}_\mathrm{H}\approx
\frac{1}{2}\,\langle\boldsymbol{\nabla\times\pi}\rangle
-\frac{1}{2}\,\boldsymbol{\nabla\cdot\pi}
&\boldsymbol{\rho}_\mathrm{H}\approx 0
\end{align}
and the transverse ($\perp$) component of the dynamical equations are
\begin{eqnarray}
\dot{\boldsymbol{E}}_\perp+3\,H\,\boldsymbol{E}_\perp&=&
\nabla^2\boldsymbol{\sigma}_\perp
+\frac{3}{2}\,(1+w)\,H^2\,\boldsymbol{\sigma}_\perp
+4\pi\,G\,(\boldsymbol{\dot{\pi}}_\perp+H\,\boldsymbol{\pi}_\perp)
\nonumber\\
\dot{\boldsymbol{\sigma}}_\perp
+2\,H\,\boldsymbol{\sigma}_\perp
&\approx&
4\pi\,G\,\boldsymbol{\pi}_\perp-\boldsymbol{E}_\perp
\end{eqnarray}
where we have used the 0th order (background space-time) expressions, $\theta=3\,H$ and 
$\rho=\frac{3\,H^2}{8\pi\,G}$, $w=p/\rho$ and $\dot{H}=-\frac{3}{2}(1+w)\,H^2$ where $H$ is the Hubble parameter.  From these we obtain a 2nd order wave equation for the tensor part of the shear
\begin{equation}
\ddot{\boldsymbol{\sigma}}_\perp
+5\,H\,\dot{\boldsymbol{\sigma}}_\perp
-\nabla^2\,\boldsymbol{\sigma}_\perp
+\frac{3}{2}\,(1-3\,w)\,H^2\,\boldsymbol{\sigma}_\perp
=8\pi\,G\,(\boldsymbol{\dot{\pi}}_\perp+2\,H\,\boldsymbol{\pi}_\perp) \ .
\end{equation}
This is consistent with the linear theory equation for the dimensionless strain which is related to the shear by
$\boldsymbol{\dot{h}}_\perp=2\,\boldsymbol{\sigma}_\perp$ as shown in \S\ref{app:TensorGeneration}.
The T part of the SVT decomposition linearized covariant perturbation theory is nearly identical as the linear tensor gauge invariant perturbation theory.  The only difference is that in covariant theory $\boldsymbol{h}_\perp$ is only defined up to transverse PSTF field which is constant in time.

\subsection{Linearized Radiation Era Tensor Shear Generation}
\label{app:TensorGeneration}

One can use the linear theory dynamics of the tensor part of the shear, $\boldsymbol{\sigma}_\perp$, to study \emph{both} the generation of gravity waves and large-scale white noise (LSWN) in kurvature.  

Using conformal time $\eta\equiv\int^t\,dt/a$, Fourier amplitudes
$\tilde\square(\eta,\boldsymbol{k})\equiv
\int d^3\boldsymbol{x}\,e^{-i\,\boldsymbol{k}\cdot\boldsymbol{x}}\,
\square(\eta,\boldsymbol{x})/(2\pi)^{3/2}$,
$\mathcal{H}(\eta)\equiv a'(\eta)/a(\eta)$

\begin{equation}
\frac{\partial^2
\boldsymbol{\tilde\sigma}_\perp(\eta,\boldsymbol{k})
}{\partial\eta^2}
+4\,\mathcal{H}(\eta)\,
\frac{\partial
\boldsymbol{\tilde\sigma}_\perp(\eta,\boldsymbol{k})
}{\partial\eta}
+
(
|\boldsymbol{k}|^2
+\frac{3}{2}\,\,(1-3\,w(\eta))\,\mathcal{H}(\eta)^2)\,
\boldsymbol{\tilde\sigma}_\perp(\eta,\boldsymbol{k})
=
\frac{8\pi\,G}{a(\eta)}\,
\frac{\partial
(a(\eta)^2\,\boldsymbol{\tilde\pi}_\perp(\eta,\boldsymbol{k}))
}{\partial\eta}
\end{equation}
which is equivalent to the well-known
\begin{equation}
\frac{\partial^2}{\partial\eta^2}
\tilde{\boldsymbol{h}}_\perp(\eta,\boldsymbol{k})+
2\,\frac{a'(\eta)}{a(\eta)}\,\frac{\partial}{\partial\eta}
\tilde{\boldsymbol{h}}_\perp(\eta,\boldsymbol{k})
+|\boldsymbol{k}|^2\,
\tilde{\boldsymbol{h}}_\perp(\eta,\boldsymbol{k})
=16\pi\,G\,a(\eta)^2\,
\tilde{\boldsymbol{\pi}}_\perp(\eta,\boldsymbol{k}) \ .
\end{equation}
where $a(\eta)$ is the scale factor and $\eta$ the conformal time.  

\end{appendix}

\printbibliography

\end{document}